\documentclass[conference]{IEEEtran}
\usepackage{epsfig}
\usepackage{url}

\title{\huge{Construction of Partial MDS (PMDS) and Sector-Disk (SD) Codes
with Two Global Parity Symbols}}

\author{\textbf{Mario Blaum}\IEEEauthorrefmark{1}, \IEEEauthorblockN{\textbf{James S. Plank}\IEEEauthorrefmark{2}, \textbf{Moshe Schwartz}\IEEEauthorrefmark{3}, and \textbf{Eitan Yaakobi}\IEEEauthorrefmark{4}\\}
\IEEEauthorblockA{\IEEEauthorrefmark{1}IBM Research Division, Almaden Research Center, San Jose, CA 95120, USA \\}
\IEEEauthorblockA{\IEEEauthorrefmark{2}Dept. of Electrical Engineering and Computer Science, Univ. of Tennessee, Knoxville, TN 37996, USA \\}
\IEEEauthorblockA{\IEEEauthorrefmark{3}Department of Electrical and Computer Engineering, Ben-Gurion University, Beer Sheva 8410501, Israel \\}
\IEEEauthorblockA{\IEEEauthorrefmark{4}Department of Electrical Engineering, California Institute of Technology, Pasadena, CA 91125, USA \\}
{\it mblaum@us.ibm.com, plank@cs.utk.edu, schwartz@ee.bgu.ac.il, yaakobi@caltech.edu\vspace{-4ex}}}
\usepackage{latexsym}
\usepackage{multirow}
\usepackage{color}
\usepackage{mathpazo}
\usepackage{times}
\usepackage{amsmath}  
\usepackage{amssymb}  
\usepackage{mathrsfs} 
 \newtheorem{theo}{Theorem}[section]
 \newtheorem{lemma}{Lemma}[section]
 
 \newtheorem{defin}{Definition}[section]
 \newtheorem{ex}{Example}[section]

\newtheorem{COROLLARY}{\indent Corollary}

\newtheorem{EXAMPLE}{\indent Example}

\newtheorem{THEOREM}{\indent Theorem}

\newtheorem{REMARK}{\indent Remark}

\newcommand{\fullstop}{\hspace{-0.85em} {\bf .}}

\newcommand{\la}{\mbox{$\leftarrow$}}

\newcommand{\al}{\mbox{$\alpha$}}

\newcommand{\eq}{\mbox{$\, =\,$}}

\newcommand{\pf}{\begin{IEEEproof}}
\newcommand{\qed}{\end{IEEEproof}}

\newcommand{\uzero}{\mbox{$\underline{0}$}}

\newcommand{\xor}{\mbox{$\,\oplus\,$}}

\newcommand{\C}{\mbox{${\cal C}$}}

\newcommand{\cO}{\mbox{${\cal O}$}}

\newcommand{\br}{\\ }
\newcommand{\ce}{\begin{center}}
\newcommand{\cen}{\end{center}}

\newcommand{\ipb}{\begin{description}}
\newcommand{\ipn}{\end{description}}

\newcommand{\qb}{\begin{quote}}
\newcommand{\qn}{\end{quote}}

\newcommand{\tp}{\begin{titlepage}}
\newcommand{\tpn}{\end{titlepage}}
\newcommand{\zb}{\begin{figure}[hbtp]}
\newcommand{\zn}{\end{figure}}

\newcommand{\EQX}[1]{\begin{equation}\label{#1}}
\newcommand{\ENX}{\end{equation}}
\newcommand{\EQL}{\begin{eqnarray*}}
\newcommand{\ENL}{\end{eqnarray*}}
\newcommand{\EQLX}[1]{\begin{eqnarray}\label{#1}}
\newcommand{\ENLX}{\end{eqnarray}}

\newcommand{\open}{\begin{document}}
\newcommand{\close}{\end{document}}
\newcommand{\lfcr}[1]{\br\hspace*{#1em}}
\newenvironment{mat}[1]
{\left[\begin{array}{#1}}{\end{array}\right]}
\newcommand{\GAMMA}{\Gamma}
\newcommand{\DELTA}{\Delta}
\newcommand{\THETA}{\Theta}
\newcommand{\LAMBDA}{\Lambda}
\newcommand{\XI}{\Xi}
\newcommand{\PI}{\Pi}
\newcommand{\SIGMA}{\Sigma}
\newcommand{\UPSILON}{\Upsilon}
\newcommand{\PHI}{\Phi}
\newcommand{\PSI}{\Psi}
\newcommand{\OMEGA}{\Omega}
\newcommand{\bldgreek}[1]{\mbox{\boldmath $#1$}}
\newcommand{\bldbeta}{\bldgreek{\beta}}
\newcommand{\bldgamma}{\bldgreek{\gamma}}
\newcommand{\blddelta}{\bldgreek{\delta}}
\newcommand{\bldepsilon}{\bldgreek{\epsilon}}
\newcommand{\bldvarepsilon}{\bldgreek{\varepsilon}}
\newcommand{\bldzeta}{\bldgreek{\zeta}}
\newcommand{\bldeta}{\bldgreek{\eta}}
\newcommand{\bldtheta}{\bldgreek{\theta}}
\newcommand{\bldvartheta}{\bldgreek{\vartheta}}
\newcommand{\bldiota}{\bldgreek{\iota}}
\newcommand{\bldkappa}{\bldgreek{\kappa}}
\newcommand{\bldlambda}{\bldgreek{\lambda}}
\newcommand{\bldmu}{\bldgreek{\mu}}
\newcommand{\bldnu}{\bldgreek{\nu}}
\newcommand{\bldxi}{\bldgreek{\xi}}
\newcommand{\bldpi}{\bldgreek{\pi}}
\newcommand{\bldvarpi}{\bldgreek{\varpi}}
\newcommand{\bldrho}{\bldgreek{\rho}}
\newcommand{\bldvarrho}{\bldgreek{\varrho}}
\newcommand{\bldsigma}{\bldgreek{\sigma}}
\newcommand{\bldvarsigma}{\bldgreek{\varsigma}}
\newcommand{\bldtau}{\bldgreek{\tau}}
\newcommand{\bldupsilon}{\bldgreek{\upsilon}}
\newcommand{\bldphi}{\bldgreek{\phi}}
\newcommand{\bldvarphi}{\bldgreek{\varphi}}
\newcommand{\bldchi}{\bldgreek{\chi}}
\newcommand{\bldpsi}{\bldgreek{\psi}}
\newcommand{\bldomega}{\bldgreek{\omega}}
\newcommand\numberthis{\addtocounter{equation}{1}\tag{\theequation}}
\renewcommand{\le}{\leqslant}
\renewcommand{\leq}{\leqslant}
\renewcommand{\ge}{\geqslant}
\renewcommand{\geq}{\geqslant}

\begin{document}
\parindent=10pt

\maketitle
\begin{abstract}
Partial MDS (PMDS) codes are erasure codes combining local (row)
correction with global additional correction of entries, while
Sector-Disk (SD) codes are erasure codes that address the mixed
failure mode of current RAID systems.
It has been an
open problem to construct general codes that have the PMDS and the SD
properties, and previous work has relied on Monte-Carlo searches.  In
this paper, we present a general construction that addresses the case
of any number of failed disks and in addition, two erased sectors. The
construction requires a modest field size.  This result generalizes
previous constructions extending RAID~5 and RAID~6.
\end{abstract}

\section{Introduction}
\label{Introduction}
Consider an $r\times n$ array whose entries are elements in a finite field
$GF(2^w)$~\cite{ms} (in general, we could consider a field $GF(p^w)$, $p$ a
prime number, but for simplicity, we constrain ourselves to binary
fields).  The array may correspond to a stripe on a disk system, where
elements co-located in the same column reside on the same disk,
or the elements may correspond to disk or SSD blocks on a large storage system.
Recent work has explored two types of erasure codes tailored for these scenarios:
Partial-MDS (PMDS) codes and Sector-Disk (SD) codes~\cite{bhh,bp}.

Both follow
the same methodology ---~$m$ entire columns of elements are devoted to coding,
and each row composes an independent~$[n,n-m,m+1]$ MDS code.
In the remaining~$n-m$ columns of the array,~$s$ more elements are also devoted to coding.
The erasure protection that they provide differentiates PMDS and SD codes.
SD codes tolerate the erasure of any~$m$ columns of elements, plus any additional~$s$
elements in the array.  PMDS codes tolerate a broader class of erasures --- any~$m$
elements per row may be erased, plus any additional~$s$ elements.

As their name implies, SD codes address the combination of disk and sector failures
that occurs in modern disk systems.  Column failures occur when entire disks break,
and sector failures can accumulate over time, typically unnoticed until an entire disk
breaks, and the failed sector is required for recovery.  PMDS codes are maximally recoverable
for codes laid out in the manner described above~\cite{bhh}.  Maximally recoverable codes have been
applied to cloud storage systems where each element resides on a different storage node~\cite{hsx}.
The rows of the array correspond to collections of storage nodes that can decode
together with good performance, while the extra~$s$ elements allow the system to tolerate broader
classes of failures.

\begin{figure}[ht]
\begin{center}
\begin{tabular}{c|c}
\epsfig{figure=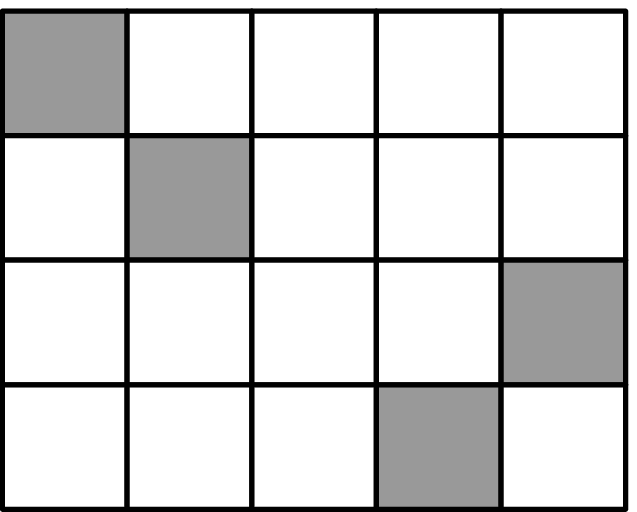,width=0.58in}
\epsfig{figure=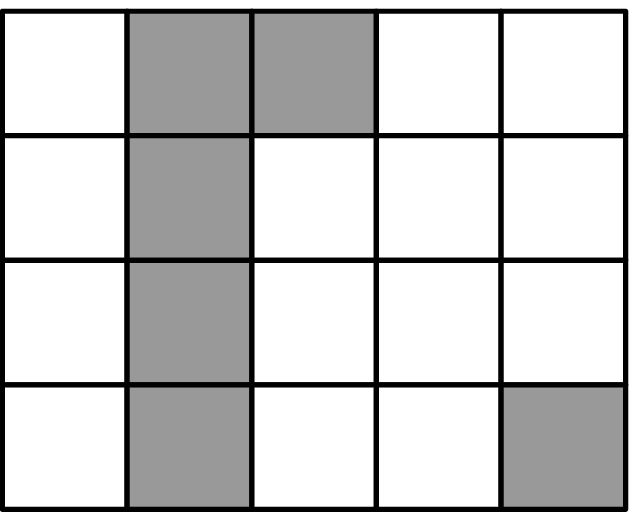,width=0.58in}
\epsfig{figure=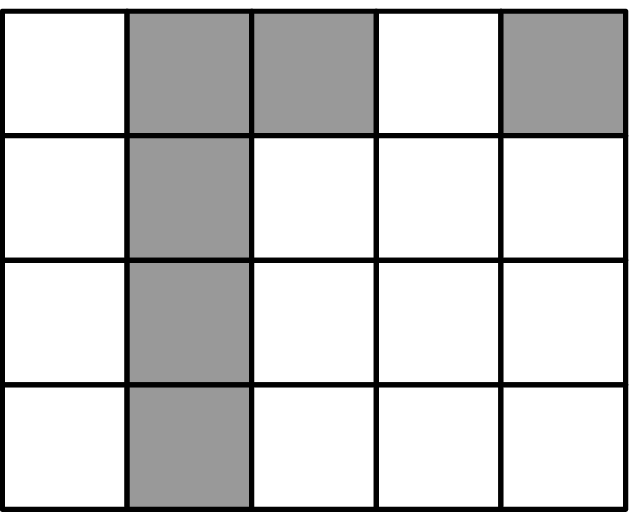,width=0.58in} &
\epsfig{figure=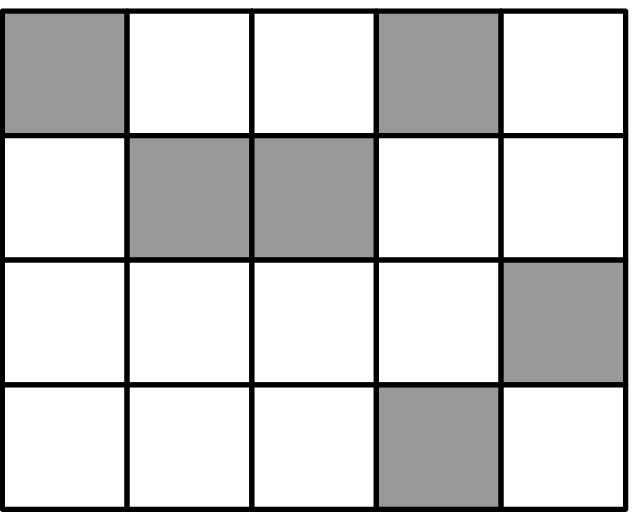,width=0.58in}
\epsfig{figure=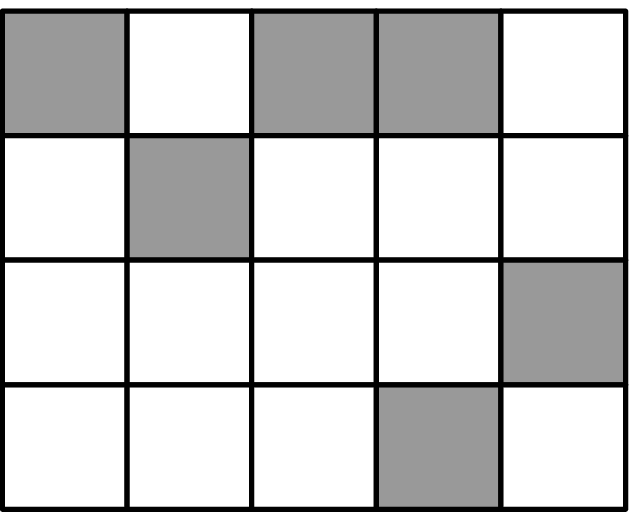,width=0.58in} \\
PMDS and SD &
PMDS only \\
\end{tabular}
\caption{Five failure scenarios on a $4\times 5$ array of elements.}
\label{figPMDS}
\end{center}
\end{figure}

We label the codes with~$(m;s)$, and illustrate the difference between PMDS and SD
in Figure~\ref{figPMDS}.  The figure depicts five failure scenarios in a $4 \times 5$ array,
encoded with a~$(1;2)$ code, where erased elements are shaded in gray.
The left scenario may be tolerated by both PMDS and SD codes,
since each row is an independent~$[5,4,2]$ MDS code.  The second two scenarios are also
tolerated
by both PMDS and SD codes, because four erasures are co-located in the same column.
The last of these is an important case, as it is not tolerated by RAID-6, even though RAID-6
devotes two full columns to coding. The two right scenarios are PMDS only, as there is one
erasure per row, plus two additional erasures.

The challenge of the current work is to define PMDS and SD codes for general parameters.
The case of $(m;1)$ PMDS codes was
solved in~\cite{bhh}. In this paper, we address the case of
$(m;2)$ PMDS and SD codes.

We begin with a formal definition of the two codes.

\begin{defin}
\label{defPMDS}
{\em
Let $\C$ be a linear $[rn,r(n-m)-s]$ code over a field such
that when codewords are taken row-wise as
$r\times n$ arrays, each row belongs in an
$[n,n-m,m+1]$ MDS code. Then,

\begin{enumerate}
\item
$\C$ is an $(m;s)$ partial-MDS (PMDS) code if,
{\em for any} $(s_1,s_2,\ldots,s_t)$ such that each $s_j\geq 1$
and $\sum_{j=1}^ts_j\eq s$, and for any $i_1,i_2,\ldots,i_t$ such that $0\leq
i_1<i_2<\cdots <i_t\leq r-1$, $\C$ can correct up
to $s_j+m$ erasures in each row $i_j$, $1\leq j\leq t$, of an array in $\C$.
\item
$\C$ is an $(m;s)$ sector-disk (SD) code if, for any
$l_1,l_2,\ldots,l_m$ such that
$0\leq l_1<l_2<\cdots <l_{m}\leq n-1$,
for any $(s_1,s_2,\ldots,s_t)$ such that each $s_j\geq 1$
and $\sum_{j=1}^ts_j\eq s$, and for any $i_1,i_2,\ldots,i_t$ such that $0\leq
i_1<i_2<\cdots <i_t\leq r-1$, $\C$ can correct up
to $s_j+m$ erasures in each row $i_j$, $1\leq j\leq t$, of an array in
$\C$ provided
that locations $l_1,l_2,\ldots l_{m}$ in each of the rows $i_j$
have been erased.
\end{enumerate}
}
\end{defin}

In the next section we give a general construction for $(m;2)$ PMDS and SD codes. Constructions of (1;2) SD codes were given in~\cite{b} and of (2;2) codes in~\cite{bp}, so our results generalize those constructions.

From now on, when we say PMDS or SD codes, we refer to $(m;2)$ PMDS or SD codes.

\section{Code Construction}
\label{construction}

Consider the field $GF(2^w)$ and let
$\al$ be an element in $GF(2^w)$.
The (multiplicative) order
of $\al$, denoted $\cO(\al)$, is the minimum $\ell>0$ such that
$\al^{\ell}\eq 1$. If $\al$ is a primitive element~\cite{ms}, then
$\cO(\al)\eq 2^w-1$. To each element $\al\in GF(2^w)$, there is an
associated (irreducible) minimal polynomial~\cite{ms} that we denote $f_{\al}(x)$.

Let $\al\in GF(2^w)$ and $rn\leq \cO(\al)$. We want to construct an
SD-code consisting of $r\times n$ arrays over $GF(2^w)$, such that $m$
of the columns correspond to parity (in RAID~5, $m\eq 1$, while in
RAID~6, $m\eq 2$). In addition, two extra symbols also correspond to
parity. When read row-wise, the codewords belong in an
$[rn,r(n-m)-2]$ code over $GF(2^w)$. Specifically,
let $\C(r,n,m,2;f_{\al}(x))$ be the $[rn,r(n-m)-2]$ code whose
$(mr+2)\times rn$ parity-check matrix is given by
\begin{eqnarray}
\label{PC}
H&=&
\left(
\begin{array}{c|c|c|c}
H_0&\uzero & \ldots &\uzero \\
\hline \uzero &H_0&\ldots &\uzero\\
\hline \vdots &\vdots &\ddots &\vdots \\
\hline \uzero &\uzero&\ldots &H_0\\
\hline H_1 & H_2 &\ldots &H_r
\end{array}
\right)
\end{eqnarray}
where
\begin{eqnarray}
\label{H0}
\hspace{-1.5ex}H_0&\hspace{-1.5ex}=\hspace{-2.5ex}&
\left(
\begin{array}{ccccc}
1 & 1 & 1 & \ldots &1\\
1 & \al & \al^2 & \ldots &\al^{n-1}\\
1 & \al^2 & \al^4 & \ldots &\al^{2(n-1)}\\
\vdots &\vdots &\vdots &\ddots &\vdots \\
1 & \al^{m-1} & \al^{2(m-1)} & \ldots &\al^{(m-1)(n-1)}\\
\end{array}\hspace{-1ex}
\right)
\end{eqnarray}
and, for $1\leq j\leq r$,
\begin{eqnarray}
\label{Hj}
\hspace{-2ex}H_j&\hspace{-2ex}=\hspace{-2.2ex}&
\left(
\begin{array}{ccccc}
\hspace{-1.5ex}1 & \hspace{-1.5ex}\al^{m} & \hspace{-1.5ex}\al^{2m} &\hspace{-2.6ex} \ldots &\hspace{-1.5ex}\al^{m(n-1)}\\
\hspace{-1.5ex}\al^{\hspace{-0.8ex}-(\hspace{-0.3ex}j-1\hspace{-0.2ex})n} &\hspace{-2.2ex} \al^{\hspace{-0.8ex}-(\hspace{-0.3ex}j-1\hspace{-0.2ex})n-1} &\hspace{-2.2ex} \al^{\hspace{-0.8ex}-(\hspace{-0.3ex}j-1\hspace{-0.2ex})n-2} &\hspace{-2.6ex} \ldots &\hspace{-1.2ex}\al^{\hspace{-0.8ex}-(\hspace{-0.3ex}j-1\hspace{-0.2ex})n-(\hspace{-0.3ex}n-1\hspace{-0.2ex})}\\
\end{array}
\hspace{-2ex}\right).
\end{eqnarray}

We will show under which conditions codes
$\C(r,n,m,2;f_{\al}(x))$ are PMDS or SD.
Unless stated otherwise, for simplicity, let us denote
$\C(r,n,m,2;f_{\al}(x))$ by $\C(r,n,m,2)$.

We start by giving some examples.
\begin{ex}
\label{excode}
{\em
Consider the finite field $GF(16)$ and let $\al$ be a primitive
element, i.e., $\cO(\al)\eq 15$. Then, the parity-check matrix of
$\C(3,5,1,2)$ is given by
\[\begin{small}
\left(
\begin{array}{ccccc|ccccc|ccccc}
1&\hspace{-2ex}1&\hspace{-2ex}1&\hspace{-2ex}1&\hspace{-2ex}1&\hspace{-0.6ex}0&\hspace{-2ex}0&\hspace{-2ex}0&\hspace{-2ex}0&\hspace{-2ex}0&\hspace{-0.6ex}0&\hspace{-2ex}0&\hspace{-2ex}0&\hspace{-2ex}0&\hspace{-2ex}0\\
\hline
0&\hspace{-2ex}0&\hspace{-2ex}0&\hspace{-2ex}0&\hspace{-2ex}0&\hspace{-0.6ex}1&\hspace{-2ex}1&\hspace{-2ex}1&\hspace{-2ex}1&\hspace{-2ex}1&\hspace{-0.6ex}0&\hspace{-2ex}0&\hspace{-2ex}0&\hspace{-2ex}0&\hspace{-2ex}0\\
\hline
0&\hspace{-2ex}0&\hspace{-2ex}0&\hspace{-2ex}0&\hspace{-2ex}0&\hspace{-0.6ex}0&\hspace{-2ex}0&\hspace{-2ex}0&\hspace{-2ex}0&\hspace{-2ex}0&\hspace{-0.6ex}1&\hspace{-2ex}1&\hspace{-2ex}1&\hspace{-2ex}1&\hspace{-2ex}1\\
\hline
1&\hspace{-2ex}\al&\hspace{-2ex}\al^2&\hspace{-2ex}\al^3&\hspace{-2ex}\al^4&\hspace{-0.6ex}1&\hspace{-2ex}\al&\hspace{-2ex}\al^2&\hspace{-2ex}\al^3& \hspace{-2ex}\al^4&\hspace{-0.6ex}1&\hspace{-2ex}\al&\hspace{-2ex}\al^2&\hspace{-2ex}\al^3&\hspace{-2ex}\al^4\\
1&\hspace{-2ex}\al^{14}&\hspace{-2ex}\al^{13}&\hspace{-2ex}\al^{12}&\hspace{-2ex}\al^{11}&
\hspace{-0.6ex}\al^{10}&\hspace{-2ex}\al^9&\hspace{-2ex}\al^8&\hspace{-2ex}\al^7&\hspace{-2ex}\al^6&\hspace{-0.6ex}\al^{5}& \hspace{-2ex}\al^{4}&\hspace{-2ex}\al^{3}&\hspace{-2ex}\al^{2}&\hspace{-2ex}\al\\
\end{array}
\right).\end{small}
\]

Similarly, the parity-check matrix of
$\C(3,5,2,2)$ is given by
\[\begin{small}
\left(
\begin{array}{ccccc|ccccc|ccccc}
1&\hspace{-2ex}1&\hspace{-2ex}1&\hspace{-2ex}1&\hspace{-2ex}1&0&\hspace{-2ex}0&\hspace{-2ex}0&\hspace{-2ex}0&\hspace{-2ex}0&0&\hspace{-2ex}0&\hspace{-2ex}0&\hspace{-2ex}0&\hspace{-2ex}0\\
1&\hspace{-2ex}\al&\hspace{-2ex}\al^2&\hspace{-2ex}\al^3&\hspace{-2ex}\al^4&0&\hspace{-2ex}0&\hspace{-2ex}0&\hspace{-2ex}0&\hspace{-2ex}0&0&\hspace{-2ex}0&\hspace{-2ex}0&\hspace{-2ex}0&\hspace{-2ex}0\\
\hline
0&\hspace{-2ex}0&\hspace{-2ex}0&\hspace{-2ex}0&\hspace{-2ex}0&1&\hspace{-2ex}1&\hspace{-2ex}1&\hspace{-2ex}1&\hspace{-2ex}1&0&\hspace{-2ex}0&\hspace{-2ex}0&\hspace{-2ex}0&\hspace{-2ex}0\\
0&\hspace{-2ex}0&\hspace{-2ex}0&\hspace{-2ex}0&\hspace{-2ex}0&1&\hspace{-2ex}\al&\hspace{-2ex}\al^2&\hspace{-2ex}\al^3&\hspace{-2ex}\al^4&0&\hspace{-2ex}0&\hspace{-2ex}0&\hspace{-2ex}0&\hspace{-2ex}0\\
\hline
0&\hspace{-2ex}0&\hspace{-2ex}0&\hspace{-2ex}0&\hspace{-2ex}0&0&\hspace{-2ex}0&\hspace{-2ex}0&\hspace{-2ex}0&\hspace{-2ex}0&1&\hspace{-2ex}1&\hspace{-2ex}1&\hspace{-2ex}1&\hspace{-2ex}1\\
0&\hspace{-2ex}0&\hspace{-2ex}0&\hspace{-2ex}0&\hspace{-2ex}0&0&\hspace{-2ex}0&\hspace{-2ex}0&\hspace{-2ex}0&\hspace{-2ex}0&1&\hspace{-2ex}\al&\hspace{-2ex}\al^2&\hspace{-2ex}\al^3&\hspace{-2ex}\al^4\\
\hline
1&\hspace{-2ex}\al^2&\hspace{-2ex}\al^4&\hspace{-2ex}\al^6&\hspace{-2ex}\al^8&1&\hspace{-2ex}\al^2&\hspace{-2ex}\al^4&\hspace{-2ex}\al^6&\hspace{-2ex}\al^8&1&\hspace{-2ex}\al^2&\hspace{-2ex}\al^4&\hspace{-2ex}\al^6&\hspace{-2ex}\al^8\\
1&\hspace{-2ex}\al^{14}&\hspace{-2ex}\al^{13}&\hspace{-2ex}\al^{12}&\hspace{-2ex}\al^{11}&
\al^{10}&\hspace{-2ex}\al^9&\hspace{-2ex}\al^8&\hspace{-2ex}\al^7&\hspace{-2ex}\al^6&\al^{5}&\hspace{-2ex}\al^{4}&\hspace{-2ex}\al^{3}&\hspace{-2ex}\al^{2}&\hspace{-2ex}\al\\
\end{array}
\right).\end{small}
\]
}
\end{ex}

Let us point out that the construction of this type of codes is valid
also over the ring of polynomials modulo $M_p(x)\eq 1+x+\cdots
+x^{p-1}$, $p$ a prime number, as done with the Blaum-Roth (BR)
codes~\cite{br}. In that case, $\cO(\al)\eq p$, where $\al^{p-1}\eq
1+\al+\cdots +\al^{p-2}$. The construction
proceeds similarly, and we denote it $\C(r,n,m,2;M_p(x))$. Utilizing
the ring modulo $M_p(x)$ allows for XOR operations at the encoding
and the decoding without look-up tables in a finite field, which is
advantageous in erasure decoding~\cite{br}. It is well known that
$M_p(x)$ is irreducible if and only if 2 is primitive in
$GF(p)$~\cite{ms}.

Next we give a lemma that is key to proving the conditions under
which codes $\C(r,n,m,2)$ are PMDS or SD.
\begin{lemma}
\label{lmain}
{\em
Let $\al\in GF(2^w)$, $rn\leq \cO(\al)$, $1\leq\ell \leq r-1$, $1\leq
s$ and, if $1\leq m\leq n-2$, let $0\leq i_0<i_1<i_2<\cdots <i_{m}\leq n-1$ and $0\leq
j_0<j_1<j_2<\cdots <j_{m}\leq n-1$. Consider the $(2m+2)\times (2m+2)$ matrix
$M(i_0,i_1,\ldots ,i_{m};j_0,j_1,\ldots ,j_{m};r;n;\ell)$ given by
\begin{eqnarray*}\begin{small}
\left(
\begin{array}{cccc|cccc}
\hspace{-2ex}1&\hspace{-2ex}1&\hspace{-2ex}\ldots &\hspace{-2ex}1&\hspace{-2ex}0&\hspace{-2ex}0&\hspace{-2ex}\ldots &\hspace{-2ex}0\hspace{0ex}\\
\hspace{-2ex}\al^{i_0}&\hspace{-2ex}\al^{i_1}&\hspace{-2ex}\ldots&\hspace{-2ex}\al^{i_m}&\hspace{-2ex}0&\hspace{-2ex}0&\hspace{-2ex}\ldots&\hspace{-2ex}0\\
\hspace{-2ex}\al^{2i_0}&\hspace{-2ex}\al^{2i_1}&\hspace{-2ex}\ldots&\hspace{-2ex}\al^{2i_m}&\hspace{-2ex}0&\hspace{-2ex}0&\hspace{-2ex}\ldots&\hspace{-2ex}0\\
\hspace{-2ex}\vdots&\hspace{-2ex}\vdots&\hspace{-2ex}\ddots&\hspace{-2ex}\vdots&\hspace{-2ex}\vdots&\hspace{-2ex}\vdots&\hspace{-2ex}\ddots&\hspace{-2ex}\vdots\\
\hspace{-2ex}\al^{(m-1)i_0}&\hspace{-2ex}\al^{(m-1)i_1}\hspace{0ex}&\hspace{-2ex}\ldots&\hspace{-2ex}\al^{(m-1)i_m}&\hspace{-2ex}0&\hspace{-2ex}0&\hspace{-2ex}\ldots&\hspace{-2ex}0\\
\hline
\hspace{-2ex}0&\hspace{-2ex}0&\hspace{-2ex}\ldots&\hspace{-2ex}0&\hspace{-2ex}1&\hspace{-2ex}1&\hspace{-2ex}\ldots&\hspace{-2ex}1\\
\hspace{-2ex}0&\hspace{-2ex}0&\hspace{-2ex}\ldots&\hspace{-2ex}0&\hspace{-2ex}\al^{j_0}&\hspace{-2ex}\al^{j_1}&\hspace{-2ex}\ldots&\hspace{-2ex}\al^{j_m}\\
\hspace{-2ex}0&\hspace{-2ex}0&\hspace{-2ex}\ldots&\hspace{-2ex}0&\hspace{-2ex}\al^{2j_0}&\hspace{-2ex}\al^{2j_1}&\hspace{-2ex}\ldots&\hspace{-2ex}\al^{2j_m}\\
\hspace{-2ex}\vdots&\hspace{-2ex}\vdots&\hspace{-2ex}\ddots&\hspace{-2ex}\vdots&\hspace{-2ex}\vdots&\hspace{-2ex}\vdots&\hspace{-2ex}\ddots&\hspace{-2ex}\vdots\\
\hspace{-2ex}0&\hspace{-2ex}0&\hspace{-2ex}0&\hspace{-2ex}\ldots&\hspace{-1ex}\al^{(m-1)j_0}&\hspace{-2ex}\al^{(m-1)j_1}&\hspace{-2ex}\ldots&\hspace{-2ex}\al^{(m-1)j_m}\hspace{-1ex}\\
\hline
\hspace{-2ex}\al^{mi_0}&\hspace{-2ex}\al^{mi_1}&\hspace{-2ex}\ldots&\hspace{-2ex}\al^{mi_m}&\hspace{-1ex}
\al^{mj_0}&\hspace{-2ex}\al^{mj_1}&\hspace{-2ex}\ldots &\hspace{-2ex}\al^{mj_m}\hspace{-1ex}\\
\hspace{-2ex}\al^{-i_0}&\hspace{-2ex}\al^{-i_1}&\hspace{-2ex}\ldots&\hspace{-2ex}\al^{-i_m}&\hspace{-1ex}
\al^{-n\ell-j_0}&\hspace{-2ex}\al^{-n\ell-j_1}&\hspace{-2ex}\ldots &\hspace{-2ex}\al^{-n\ell-j_m}\hspace{-1ex}\\
\end{array}\hspace{-1.5ex}
\right)
\end{small}
\end{eqnarray*}

Let
\begin{multline*}
\Delta(i_0,i_1,\ldots ,i_{m};j_0,j_1,\ldots ,j_{m};r;n;\ell) \\
 \eq\det M(i_0,i_1,\ldots ,i_{m};j_0,j_1,\ldots ,j_{m};r;n;\ell).
\end{multline*}
Then,
\begin{align*}
&\Delta(i_0,i_1,\ldots ,i_{m};j_0,j_1,\ldots ,j_{m};r;n;s;\ell)& & & & &
\end{align*}
\begin{equation}\label{eqmain}
\hspace{-1ex}=\hspace{-1ex}\left(\prod_{0\leq u<v\leq m}\hspace{-3ex}\left(\hspace{-0.5ex}\al^{i_u}\hspace{-0.5ex}\xor\hspace{-0.5ex}\al^{i_v}\hspace{-0.5ex}\right)\hspace{-1ex}\left(\hspace{-0.5ex}\al^{j_u}\hspace{-0.5ex}\xor\hspace{-0.5ex}\al^{j_v}\hspace{-0.5ex}\right)\hspace{-1ex}
\right)\hspace{-1ex}\left(\hspace{-0.5ex}\al^{-\sum_{u=0}^mi_u}\hspace{-0.5ex}\xor\hspace{-0.5ex}\al^{-n\ell-\sum_{u=0}^mj_u}\hspace{-0.5ex}\right)\hspace{-0.5ex}.
\end{equation}}
\end{lemma}
\pf
For simplicity, let us denote \[\Delta\eq \Delta(i_0,i_1,\ldots ,i_{m};j_0,j_1,\ldots ,j_{m};r;n;\ell).\]
Consider the $m\times (m+1)$ matrices
\[M=
\left(
\begin{array}{cccc}
1&1&\ldots &1\\
\al^{i_0}&\al^{i_1}&\ldots &\al^{i_m}\\
\al^{2i_0}&\al^{2i_1}&\ldots &\al^{2i_m}\\
\vdots&\vdots&\ddots &\vdots\\
\al^{(m-1)i_0}&\al^{(m-1)i_1}&\ldots &\al^{(m-1)i_m}\\
\end{array}
\right)\]
and
\[M'=
\left(
\begin{array}{cccc}
1&1&\ldots &1\\
\al^{j_0}&\al^{j_1}&\ldots &\al^{j_m}\\
\al^{2j_0}&\al^{2j_1}&\ldots &\al^{2j_m}\\
\vdots&\vdots&\ddots &\vdots\\
\al^{(m-1)j_0}&\al^{(m-1)j_1}&\ldots &\al^{(m-1)j_m}\\
\end{array}
\right).
\]

For each $u$, $0\leq u\leq m$, let $M_u$ and $M'_u$ denote the
$m\times m$ Vandermonde matrices obtained from deleting column $u$
from $M$ and $M'$ respectively. Also, for $0\leq u,v\leq 2m+1$,
$u\neq v$, let $X^{(u,v)}$ be the $(2m)\times (2m)$ matrix obtained from removing
columns $u$ and $v$ and the last two rows from $M(i_0,i_1,\ldots
,i_{m};j_0,j_1,\ldots ,j_{m};r;n;\ell)$.

If $0\leq u,v\leq m$, $u\neq v$,
\[
X^{(u,v)}=\left(
\begin{array}{c|c}
P&\uzero\\
\hline
\uzero & M'
\end{array}
\right),
\]
where $P$ denotes an $m\times(m-1)$ matrix and $\uzero$ are zero
matrices. Notice that $X^{(u,v)}$ has rank smaller than $2m$, since
the first $m$ rows have rank smaller than $m$. Thus,
\[
\det\left(X^{(u,v)}\right)=0\quad\text{for $0\leq u,v\leq m$, $u\neq v$.}
\]
If $0\leq u\leq m$ and $m+1\leq v\leq 2m+1$,
\[
X^{(u,v)}=\left(
\begin{array}{c|c}
M_u&\uzero\\
\hline
\uzero & M'_{v-m-1}
\end{array}
\right).
\]

By properties of determinants,
\[
\det\left(X^{(u,v)}\right)=\left(\det(M_u)\right)\left(\det(M'_{v-m-1})\right)
\]
for $0\leq u\leq m$, $m+1\leq v\leq 2m+1$.
Similarly,
\[
\det\left(X^{(u,v)}\right)=\left(\det(M'_{u-m-1})\right)\left(\det(M_v)\right)
\]
for $m+1\leq u\leq 2m+1$, $0\leq v\leq m$, and
\[
\det\left(X^{(u,v)}\right)=0,
\]
for $m+1\leq u,v\leq 2m+1$, $u\neq v$.

Expanding the determinant $\Delta$ from the bottom row, and then from the next to bottom row, we obtain
\begin{align}
\Delta=&\left(\bigoplus_{u=0}^m\,\al^{-i_u}\bigoplus_{v=0\atop v\neq u}^{m}\,\al^{mi_v}\det \left(X^{(u,v)}\right)\right) & \nonumber \\
& \xor \left(\bigoplus_{u=0}^m\,\al^{-i_u}\bigoplus_{v=m+1}^{2m+1}\,\al^{m{j_{v-m-1}}}\det \left(X^{(u,v)}\right)\right) & \nonumber \\
& \xor \left(\bigoplus_{u=m+1}^{2m+1}\,\al^{-n\ell-j_{u-m-1}}\bigoplus_{v=0\atop v\neq u}^{m}\,\al^{mi_v}\det \left(X^{(u,v)}\right)\right)& \nonumber \\
& \xor\left(\bigoplus_{u=m+1}^{2m+1}\hspace{-1ex}\al^{-n\ell-j_{u-m-1}} \hspace{-1.5ex}\bigoplus_{v=m+1}^{2m+1}\hspace{-1ex}\al^{m{j_{v-m-1}}}\det \left(X^{(u,v)}\right)\hspace{0ex}\right)& \nonumber \\
=& \left(\bigoplus_{u=0}^m\,\al^{-i_u}\bigoplus_{v=0}^{m}\,\al^{mj_v}\det (M_u)\det (M'_v)\right) & \nonumber \\
& \xor \left(\bigoplus_{u=0}^m\,\al^{-n\ell -j_u}\bigoplus_{v=0}^{m}\,\al^{mi_v}\det (M_v)\det (M'_u)\right) &\nonumber \\
=&\left(\bigoplus_{u=0}^m\,\al^{-i_u}\det (M_u)\right)\left(\bigoplus_{u=0}^{m}\,\al^{mj_u}\det (M'_u)\right)&\nonumber \\
\label{ddelta}
& \xor \left(\bigoplus_{u=0}^m\,\al^{-n\ell -j_u}\det (M'_u)\right)\left(\bigoplus_{u=0}^{m}\,\al^{mi_u}\det (M_u)\right).
\end{align}

Let
\begin{eqnarray*}
W_0&=&
\left(
\begin{array}{cccc}
1&1&\ldots &1\\
\al^{i_0}&\al^{i_1}&\ldots &\al^{i_m}\\
\al^{2i_0}&\al^{2i_1}&\ldots &\al^{2i_m}\\
\vdots&\vdots&\ddots &\vdots\\
\al^{(m-1)i_0}&\al^{(m-1)i_1}&\ldots&\al^{(m-1)i_m}\\
\al^{mi_0}&\al^{mi_1}&\ldots &\al^{mi_m}\\
\end{array}
\right)\\
W_1&=&
\left(
\begin{array}{cccc}
1&1&\ldots &1\\
\al^{i_0}&\al^{i_1}&\ldots &\al^{i_m}\\
\al^{2i_0}&\al^{2i_1}&\ldots &\al^{2i_m}\\
\vdots&\vdots&\ddots &\vdots\\
\al^{(m-1)i_0}&\al^{(m-1)i_1}&\ldots&\al^{(m-1)i_m}\\
\al^{-i_0}&\al^{-i_1}&\ldots &\al^{-i_m}\\
\end{array}
\right)
\end{eqnarray*}
\begin{eqnarray*}
W'_0&=&
\left(
\begin{array}{cccc}
1&1&\ldots &1\\
\al^{j_0}&\al^{j_1}&\ldots &\al^{j_m}\\
\al^{2j_0}&\al^{2j_1}&\ldots &\al^{2j_m}\\
\vdots&\vdots&\ddots &\vdots\\
\al^{(m-1)j_0}&\al^{(m-1)j_1}&\ldots&\al^{(m-1)j_m}\\
\al^{mj_0}&\al^{mj_1}&\ldots &\al^{mj_m}\\
\end{array}
\right)\\
W'_1&=&
\left(
\begin{array}{cccc}
1&1&\ldots &1\\
\al^{j_0}&\al^{j_1}&\ldots &\al^{j_m}\\
\al^{2j_0}&\al^{2j_1}&\ldots &\al^{2j_m}\\
\vdots&\vdots&\ddots &\vdots\\
\al^{(m-1)j_0}&\al^{(m-1)j_1}&\ldots&\al^{(m-1)j_m}\\
\al^{-m\ell-j_0}&\al^{-m\ell-j_1}&\ldots &\al^{-m\ell-j_m}
\end{array}
\right)
\end{eqnarray*}

Notice that, by properties of determinants and of Vandermonde determinants,
\begin{small}
\begin{align*}
&\det (W_0)=\bigoplus_{u=0}^m\al^{mi_u}\det (M_u)=\hspace{-1ex}\prod_{0\leq u<v\leq m}\hspace{-2ex}(\al^{i_u}\xor \al^{i_v})\\
&\det (W_1)=\bigoplus_{u=0}^m\al^{-i_u}\det (M_u)=\al^{-\sum_{u=0}^mi_u}\hspace{-2ex}\prod_{0\leq u<v\leq m}\hspace{-2ex}(\al^{i_u}\xor \al^{i_v})\\
&\det (W'_0)=\bigoplus_{u=0}^m\al^{mj_u}\det (M'_u)=\hspace{-1ex}\prod_{0\leq u<v\leq m}\hspace{-2ex}(\al^{j_u}\xor \al^{j_v})\\
&\det (W'_1)=\bigoplus_{u=0}^m\al^{-m\ell-j_u}\det (M'_u)\hspace{-0.5ex}=\hspace{-0.5ex}\al^{-m\ell-\sum_{u=0}^m\hspace{-0.5ex}j_u}\hspace{-3ex}\prod_{0\leq u<v\leq m}\hspace{-2.5ex}(\al^{j_u}\hspace{-0.5ex}\xor \al^{j_v}\hspace{-0.2ex}).
\end{align*}\end{small}
So, (\ref{ddelta}) becomes
\begin{align*}
 \Delta=&
\left(
\begin{array}{cc}
\det (W_0)&\det(W'_0)\\
\det (W_1)&\det(W'_1)\\
\end{array}
\right) & \\
=&\left(\prod_{0\leq u<v\leq m}\left(\al^{i_u}\xor
\al^{i_v}\right)\left(\al^{j_u}\xor \al^{j_v}\right)\right) & \\
\cdot&\det\left(
\begin{array}{cc}
1&1\\
\al^{-\sum_{u=0}^mi_u}&\al^{-m\ell-\sum_{u=0}^mj_u}\\
\end{array}
\right) &
\end{align*}
and (\ref{eqmain}) follows.
\qed

Lemma~\ref{lmain} is valid also over the ring of polynomials modulo
$M_p(x)$, $p$ prime, where $rn< p$. Let us illustrate it with an
example for $m\eq 1$ and $m\eq 2$.

\begin{ex}
\label{exm1m2}
{\em
Let $m\eq 1$, then
\begin{align*}
& M(i_0,i_1;j_0,j_1;r;n;\ell) & \\
& = \left(
\begin{array}{cc|cc}
1&1&0&0\\
\hline
0&0&1&1\\
\hline
\al^{i_0}&\al^{i_1}&\al^{j_0}&\al^{j_1}\\
\al^{-i_0}&\al^{-i_1}&\al^{-n\ell-j_0}&\al^{-n\ell-j_1}\\
\end{array}
\right) &
\end{align*}
and
\begin{align*}
& \Delta(i_0,i_1;j_0,j_1;r;n;s;\ell) &\\
& =\left(\al^{i_0}\xor\al^{i_1}\right)\left(\al^{j_0}\xor\al^{j_1}\right)\left(\al^{-i_0-i_1}\xor\al^{-n\ell-j_0-j_1}\right). &
\end{align*}

If $m\eq 2$, Lemma~\ref{lmain} gives
\begin{align*}
& M(i_0,i_1,i_2;j_0,j_1,j_2;r;n;\ell)& \\
&=
\left(
\begin{array}{ccc|ccc}
1&\hspace{-2ex}1&\hspace{-2ex}1&0&\hspace{-2ex}0&\hspace{-2ex}0\\
\al^{i_0}&\hspace{-2ex}\al^{i_1}&\hspace{-2ex}\al^{i_2}&0&\hspace{-2ex}0&\hspace{-2ex}0\\
\hline
0&\hspace{-2ex}0&\hspace{-2ex}0&1&\hspace{-2ex}1&\hspace{-2ex}1\\
0&\hspace{-2ex}0&\hspace{-2ex}0&\al^{j_0}&\hspace{-2ex}\al^{j_1}&\hspace{-2ex}\al^{j_2}\\
\hline
\al^{2i_0}&\hspace{-2ex}\al^{2i_1}&\hspace{-2ex}\al^{2i_2}&\al^{j_0}&\hspace{-2ex}\al^{j_1}&\hspace{-2ex}\al^{j_2}\\
\al^{-i_0}&\hspace{-2ex}\al^{-i_1}&\hspace{-2ex}\al^{-i_2}&\al^{-n\ell-j_0}&\hspace{-2ex}\al^{-n\ell-j_1}&\hspace{-2ex}\al^{-n\ell-j_2}\\
\end{array}
\right) &
\end{align*}
and
\begin{align*}
& \Delta(i_0,i_1,i_2;j_0,j_1,j_2;r;n;s;\ell) & \\
& = \left(\al^{i_0}\xor\al^{i_1}\right)\left(\al^{i_0}\xor\al^{i_2}\right) \left(\al^{i_1}\xor\al^{i_2}\right)\left(\al^{j_0}\xor\al^{j_1}\right) & \\
& \left(\al^{j_0}\xor\al^{j_2}\right)\hspace{-0.5ex}\left(\al^{j_1}\xor\al^{j_2}\right)\hspace{-0.5ex} \left(\al^{-i_0-i_1-i_2}\xor\al^{-n\ell-j_0-j_1-j_2}\right)\hspace{-0.5ex}. &
\end{align*}
}
\end{ex}

In the next section we study codes $\C(r,n,m,2;f_{\al}(x))$ and $\C(r,n,m,2;M_p(x))$ as SD and PMDS codes.

\section{Construction of SD and PMDS codes}
\label{SDandPMDS}

Let us start with our main result for SD codes.
\begin{theo}
\label{theoSD}
{\em
The codes $\C(r,n,m,2;f_{\al}(x))$ and $\C(r,n,m,2;M_p(x))$ are SD.
}
\end{theo}
\pf Assume that $m$ columns have been erased and in addition we have
two random erasures. Assume first that these two random erasures occurred in the same
row $\ell$ of the stripe. The rows that are different from $\ell$ are
corrected since each one of them has $m$ erasures, which are handled
by the horizontal code, that is, each horizontal code is given by the
parity-check matrix $H_0$, which is the parity-check matrix of a RS
code that can correct up to $m$ erasures~\cite{ms}. Thus, we have to
solve a linear system with
$m+2$ unknowns. Without loss of generality, assume that the erasures
in row $\ell$ occurred in locations
$i_0,i_1,\ldots,i_m,i_{m+1}$, where $0\leq i_0<i_1<\cdots
<i_m<i_{m+1}\leq n$. According to the parity-check matrix of the code as
given by~(\ref{PC}), (\ref{H0}), and~(\ref{Hj}), there will be a unique
solution if and only if the $(m+2)\times (m+2)$ matrix
\[\left(
\begin{array}{ccccc}
1&1&\ldots &1&1\\
\al^{i_0}&\al^{i_1}&\ldots &\al^{i_m}&\al^{i_{m+1}}\\
\al^{2i_0}&\al^{2i_1}&\ldots &\al^{2i_m}&\al^{2i_{m+1}}\\
\vdots &\vdots &\ddots &\vdots &\vdots \\
\al^{mi_0}&\al^{mi_1}&\ldots &\al^{mi_m}&\al^{mi_{m+1}}\\
\al^{-n\ell -i_0}&\al^{-n\ell -i_1}&\ldots &\al^{-n\ell
-i_m}&\al^{-n\ell -i_{m+1}}\\
\end{array}
\right)\]
is invertible. By taking $\al^{-n\ell}$ in the last row as a common
factor, and by multiplying each column $j$, $0\leq j\leq m+1$, by
$\al^{i_j}$, this matrix is transformed into a Vandermonde matrix,
which is always invertible in a field and also in the ring of
polynomials modulo $M_p(x)$~\cite{br}.

Consider now the case in which the two random failures occur in
different rows. Specifically, assume that columns
$i_0,i_1,\ldots,i_{m-1}$ were erased, where
$0\leq i_0<i_1<\ldots <i_{m-1}\leq n-1$, and in addition, entries
$(\ell,t)$ and $(\ell',t')$ were erased, where
$t,t'\not\in\{i_0,i_1,\ldots,i_{m-1}\}$ and $0\leq\ell <\ell'\leq r-1$.
Again, using the parity-check matrix of the code as
given by~(\ref{PC}), (\ref{H0}), and~(\ref{Hj}), there will be a unique
solution if and only if the $(2m+2)\times (2m+2)$ matrix

\vspace{-3.5ex}
\begin{small}\begin{align*}
\hspace{-1ex}\left(
\begin{array}{cccc|cccc}
\hspace{-2ex}1&\hspace{-2ex}1&\hspace{-2ex}\ldots &\hspace{-2ex}1&\hspace{-1ex}0&\hspace{-2ex}0&\hspace{-2ex}\ldots &\hspace{-2ex}0\hspace{-2ex}\\
\hspace{-2ex}\al^{i_0}&\hspace{-2ex}\al^{i_1}&\hspace{-2ex}\ldots&\hspace{-2ex}\al^{t}&\hspace{-1ex}0&\hspace{-2ex}0&\hspace{-2ex}\ldots&\hspace{-2ex}0\hspace{-2ex}\\
\hspace{-2ex}\al^{2i_0}&\hspace{-2ex}\al^{2i_1}&\hspace{-2ex}\ldots&\hspace{-2ex}\al^{2t}&\hspace{-1ex}0&\hspace{-2ex}0&\hspace{-2ex}\ldots&\hspace{-2ex}0\hspace{-2ex}\\
\hspace{-2ex}\vdots&\hspace{-2ex}\vdots&\hspace{-2ex}\ddots &\hspace{-2ex}\vdots&\hspace{-1ex}\vdots&\hspace{-2ex}\vdots&\hspace{-2ex}\ddots&\hspace{-2ex}\vdots\hspace{-2ex}\\
\hspace{-2ex}\al^{(m-1)i_0}&\hspace{-2ex}\al^{(m-1)i_1}&\hspace{-2ex}\ldots&\hspace{-2ex}\al^{(m-1)t}&\hspace{-1ex}0&\hspace{-2ex}0&\hspace{-2ex}\ldots&\hspace{-2ex}0\hspace{-2ex}\\
\hline
\hspace{-2ex}0&\hspace{-2ex}0&\hspace{-2ex}\ldots &\hspace{-2ex}0&\hspace{-1ex}1&\hspace{-2ex}1&\hspace{-2ex}\ldots &\hspace{-2ex}1\hspace{-2ex}\\
\hspace{-2ex}0&\hspace{-2ex}0&\hspace{-2ex}\ldots &\hspace{-2ex}0&\hspace{-1ex}\al^{i_0}&\hspace{-2ex}\al^{i_1}&\hspace{-2ex}\ldots &\hspace{-2ex}\al^{t'}\hspace{-2ex}\\
\hspace{-2ex}0&\hspace{-2ex}0&\hspace{-2ex}\ldots&\hspace{-2ex}0&\hspace{-1ex}\al^{2i_0}&\hspace{-2ex}\al^{2i_1}&\hspace{-2ex}\ldots&\hspace{-2ex}\al^{2t'}\hspace{-2ex}\\
\hspace{-2ex}\vdots&\hspace{-2ex}\vdots&\hspace{-2ex}\ddots&\hspace{-2ex}\vdots&\hspace{-1ex}\vdots&\hspace{-2ex}\vdots&\hspace{-2ex}\ddots&\hspace{-2ex}\vdots\hspace{-2ex}\\
\hspace{-2ex}0&\hspace{-2ex}0&\hspace{-2ex}\ldots &\hspace{-2ex}0&\hspace{-1ex}\al^{(m-1)i_0}&\hspace{-2ex}\al^{(m-1)i_1}&\hspace{-2ex}\ldots&\hspace{-2ex}\al^{(m-1)t'}\hspace{-2ex}\\
\hline
\hspace{-2ex}\al^{mi_0}&\hspace{-2ex}\al^{mi_1}&\hspace{-2ex}\ldots &\hspace{-2ex}\al^{mt}&\hspace{-1ex}\al^{mi_0}&\hspace{-2ex}\al^{mi_1}&\hspace{-2ex}\ldots &\hspace{-2ex}\al^{mt'}\hspace{-2ex}\\
\hspace{-2ex}\al^{-n\ell-i_0}&\hspace{-2ex}\al^{-n\ell-i_1}&\hspace{-2ex}\ldots &\hspace{-2ex}\al^{-n\ell-t}&\hspace{-1ex}\al^{-n\ell'-i_0}&\hspace{-2ex}\al^{-n\ell'-i_1}&\hspace{-2ex}\ldots &\hspace{-2ex}\al^{-n\ell'-t'}\hspace{-2ex}
\end{array}
\right)\vspace{-2.3ex} 
\end{align*}\end{small}
\hspace{-2.3ex}
is invertible. Taking $\al^{-n\ell}$ as a common factor in the last row, we
obtain the matrix
\[M(i_0,i_1,i_2,\ldots ,i_{m-1},t;i_0,i_1,i_2,\ldots ,i_{m-1},t';r;n;\ell'-\ell)\]
as defined in Lemma~\ref{lmain}, whose determinant, by~(\ref{eqmain}), is
given by
\begin{align*}
&  \Delta(i_0,i_1,i_2,\ldots ,i_{m-1},t;i_0,i_1,i_2,\ldots ,i_{m-1},t';r;n;\ell'\hspace{-0.3ex} -\hspace{-0.3ex} \ell) \\
& \quad =\left(\prod_{0\leq u<v\leq m-1}\left(\al^{i_u}\xor\al^{i_v}\right)^2\right) \\
&\quad\quad\ \cdot \left(\prod_{0\leq u\leq m-1}\left(\al^{i_u}\xor\al^{t}\right)\left(\al^{i_u}\xor\al^{t'}\right)\right) \\
&\quad\quad\ \cdot \al^{-\sum_{u=0}^{m-1}i_u}\left(\al^{-t}\xor\al^{-n(\ell'-\ell)-t'} \right).
\end{align*}
For simplicity, redefine $\ell\la \ell'-\ell$, hence, $1\leq\ell\leq r-1$.
Each binomial $\left(\al^{i_u}\xor\al^{i_v}\right)$,
$\left(\al^{i_u}\xor\al^{t}\right)$ and
$\left(\al^{i_u}\xor\al^{t'}\right)$ above is invertible, so it
remains to be proven that $\left(\al^{-t}\xor\al^{-n\ell-t'}\right)$
is invertible. If it is not, $n\ell+t'-t\equiv 0\;(\bmod \cO(\al))$.
But
\begin{multline*}
 0<n\ell+t'-t\leq n(r-1)+t'-t \\
 \eq nr-\left(n-(t'-t)\right)\leq nr-1<\cO(\al),
\end{multline*}
so, $n\ell+t'-t\not\equiv 0\;(\bmod \cO(\al)$.
\qed

Next, let us prove a similar result for PMDS codes. In fact, codes
$\C(r,n,m,2;f_{\al}(x))$ and $\C(r,n,m,2;M_p(x))$ are not PMDS, but
we will obtain PMDS codes
with a modification that requires a larger field or
ring. Let
\begin{eqnarray}
\label{Nnm}
N&=&(m+1)(n-m-1)+1
\end{eqnarray}
$\al\in GF(2^w)$ and $rN \leq \cO(\al)$.
As in the case of SD codes, we construct a
PMDS code consisting of $r\times n$ arrays over $GF(2^w)$, such that $m$
of the columns correspond to parity and in addition, two extra
symbols also correspond to
parity. When read row-wise, the codewords belong in an
$[rn,r(n-m)-2]$ code over $GF(2^w)$. Specifically,
let $\C'(r,n,m,2;f_{\al}(x))$ be the $[rn,r(n-m)-2]$ code whose
$(mr+2)\times rn$ parity-check matrix is given by~$H'$, which is
identical to~$H$ in (\ref{PC}), except the bottommost two
rows are defined as:
\[
\left(
\begin{array}{c|c|c|c}
H'_1 & H'_2 &\ldots &H'_r
\end{array}
\right) \]
where, for $1\leq j\leq r$,
\begin{small}
\begin{eqnarray*}\label{Hj'}
\hspace{-0.7ex}H'_j&\hspace{-2ex}=\hspace{-2.3ex}&
\left(\hspace{-0.3ex}
\begin{array}{ccccc}
1 & \al^{m} & \al^{2m} &\hspace{-3ex} \ldots &\al^{m(n-1)}\\
\hspace{-1.5ex}\al^{-(j-1)N} & \hspace{-2ex}\al^{-(j-1)N-1} & \hspace{-2ex}\al^{-(j-1)N-2} & \hspace{-3ex}\ldots &\hspace{-2ex}\al^{-(j-1)N-(n-1)}\hspace{-2ex}\\
\end{array}
\right)\hspace{-0.7ex}.
\end{eqnarray*}
\end{small}

As before, the construction is also valid over the ring of
polynomials $M_p(x)$, $p$ prime, in which case we denote the codes
$\C'(r,n,m,2;M_p(x))$.
Let us give an example.
\begin{ex}
\label{exC'}
{\em
Let $n\eq 5$, $m\eq 1$ and $r\eq 3$. According to~(\ref{Nnm}), $N\eq
(2)(3)+1\eq 7$. Thus, we need $\cO(\al)\,>\, rN\eq 21$. For instance
we may consider the field $GF(32)$ and $\al$ primitive in
$GF(32)$, i.e., $\cO(\al)\eq 31>21$ (we can also handle
$r\eq 4$ in this example). Thus,
the parity-check matrix of
$\C'(3,5,1,2;f_{\al}(x))$ is given by

\vspace{-2ex}
\begin{small}
\[
\left(
\begin{array}{ccccc|ccccc|ccccc}
\hspace{-1.5ex}1&\hspace{-2ex}1&\hspace{-2ex}1&\hspace{-2ex}1&\hspace{-2ex}1&\hspace{-1ex}0&\hspace{-2ex}0&\hspace{-2ex}0&\hspace{-2ex}0&\hspace{-2ex}0&\hspace{-1ex}0&\hspace{-2ex}0&\hspace{-2ex}0&\hspace{-2ex}0&\hspace{-2ex}0\hspace{-2ex}\\
\hline
\hspace{-1.5ex}0&\hspace{-2ex}0&\hspace{-2ex}0&\hspace{-2ex}0&\hspace{-2ex}0&\hspace{-1ex}1&\hspace{-2ex}1&\hspace{-2ex}1&\hspace{-2ex}1&\hspace{-2ex}1&\hspace{-1ex}0&\hspace{-2ex}0&\hspace{-2ex}0&\hspace{-2ex}0&\hspace{-2ex}0\hspace{-2ex}\\
\hline
\hspace{-1.5ex}0&\hspace{-2ex}0&\hspace{-2ex}0&\hspace{-2ex}0&\hspace{-2ex}0&\hspace{-1ex}0&\hspace{-2ex}0&\hspace{-2ex}0&\hspace{-2ex}0&\hspace{-2ex}0&\hspace{-1ex}1&\hspace{-2ex}1&\hspace{-2ex}1&\hspace{-2ex}1&\hspace{-2ex}1\hspace{-2ex}\\
\hline
\hspace{-1.5ex}1&\hspace{-2ex}\al&\hspace{-2ex}\al^2&\hspace{-2ex}\al^3&\hspace{-2ex}\al^4&\hspace{-1ex}1&\hspace{-2ex}\al&\hspace{-2ex}\al^2&\hspace{-2ex}\al^3&\hspace{-2ex}\al^4&\hspace{-1ex}1&\hspace{-2ex}\al&\hspace{-2ex}\al^2&\hspace{-2ex}\al^3&\hspace{-2ex}\al^4\hspace{-2ex}\\
\hline
\hspace{-1.5ex}1&\hspace{-2ex}\al^{30}&\hspace{-2ex}\al^{29}&\hspace{-2ex}\al^{28}&\hspace{-2ex}\al^{27}&\hspace{-1ex}\al^{24}&\hspace{-2ex}\al^{23}&\hspace{-2ex}\al^{22}&\hspace{-2ex}\al^{21}&\hspace{-2ex}\al^{20}&\hspace{-1ex}\al^{17}&\hspace{-2ex}\al^{16}&\hspace{-2ex}\al^{15}&\hspace{-2ex}\al^{14}&\hspace{-2ex}\al^{13}\hspace{-2ex}\\
\end{array}
\right).
\]
\end{small}
}
\end{ex}

\begin{theo}
\label{theoPMDS}
{\em
The codes $\C'(r,n,m,2;f_{\al}(x))$ and $\C'(r,n,m,2;M_p(x))$ are PMDS.
}
\end{theo}

Theorem~\ref{theoPMDS} is proven similarly to Theorem~\ref{theoSD}.
For reasons of space, we omit the proof here.


\section{Conclusion}

We have described a construction for SD codes and PMDS codes where
the number of additional sectors,~$s$ equals two.
The minimal field size required by
the construction for SD codes is only the total number of sectors in the
array, and
in the case of PMDS codes, at most of quadratic order in the total
number of sectors.

Further results, not described in detail due to the space limit, are a
construction for $(m;s)$ SD codes restricted to not having two of the
$s$ erased sectors in the same disk, and a construction for
$(m;s)$-like PMDS codes with extra redundancy of order $O(s\log s)$.

As related work, let us mention
a recent paper~\cite{ghjy} that gives constructions of $(1;s)$ PMDS
codes trying to minimize the size of the field. In fact, $(1;s)$ PMDS,
called Maximally Recoverable codes in~\cite{ghjy}, satisfy also the
requirements of Locally Repairable codes~\cite{pd,tb}.
Additionally, the recently-defined STAIR codes relax the failure-coverage
of SD codes in order to allow for general constructions~\cite{ll}.

%
\section*{Acknowledgment}
This work is supported by the National Science Foundation,
under grant CSR-1016636, and by an IBM Faculty Award.

\end{document}